\documentstyle[prl,aps,twocolumn]{revtex}
\begin{document}
\draft

\twocolumn[\hsize\textwidth\columnwidth\hsize\csname @twocolumnfalse\endcsname
\title{Fundamental Constants and Conservation Laws}
\author{Heui-Seol Roh\thanks{e-mail: hroh@nature.skku.ac.kr}}
\address{BK21 Physics Research Division, Department of Physics, Sung Kyun Kwan University, Suwon 440-746, Republic of Korea}
\date{\today}
\maketitle

\begin{abstract}
This work describes underlying features of the universe such as fundamental constants
and cosmological parameters, conservation laws, baryon and lepton asymmetries, etc. in
the context of local gauge theories for fundamental forces under the constraint of the
flat universe. Conservation laws for fundamental forces are related to gauge theories
for fundamental forces, their resulting fundamental constants are quantitatively
analyzed, and their possible violations at different energy scales are proposed based
on experimental evidences.
\end{abstract}

\pacs{PACS numbers:  11.30.-j, 11.30.Er, 11.30.Fs}
]
\narrowtext

One of the major developments in modern physics is the understanding of fundamental
constants and conservation laws for fundamental forces governing the universe.
Although some of them are clear, there still exist some mysterious fundamental
constants and conservation laws in nature, which make underlying principles of the
universe be complicated. The origin of several fundamental constants are, on the one
hand, not clarified: for example, Newton gravitation constant $G_N \simeq 10^{-38} \
\textup{GeV}^{-2}$, cosmological constant $\Lambda_0 \simeq 10^{-84} \
\textup{GeV}^2$, Hubble constant $H_0 \simeq 10^{-42}$ GeV, baryon asymmetry $\delta_B
\simeq 10^{-10}$, and $\Theta \leq 10^{-9}$ \cite{Eins0,Hubb,Jaff,Stei0,Alta}. The
origin of several conservation laws are, on the other hand, not understandable:
conservation laws of proton number, baryon number, and lepton number and violations of
discrete symmetries (P, C, T, CP). The mysteries of fundamental constants may be
uncovered if their origins can be traced by linking fundamental conservation laws to
local gauge theories possessing symmetries. Quantum gauge theories for fundamental
forces hold commonly underlying principles such as special relativity, quantum
mechanics, and gauge invariance, and fundamental constants such as the Planck constant
($h$) and the light velocity ($c$) come from underlying principles, quantum mechanics
and special relativity. In this context, the other fundamental constants and
cosmological constants encountered in physics are considered in depth with relations
to conservation laws associated with fundamental forces as the consequence of gauge
invariance: other challenging problems including the baryon asymmetry and lepton
asymmetry may be systematically investigated as well. This paper thus intends to
report brief summaries for fundamental constants and conservation laws, whose details
are discussed in references \cite{Roh1,Roh11,Roh2,Roh3,Roh31}. The above references
deal with substantial steps toward the unification of fundamental forces beyond
Einstein's general relativity for gravitational interactions \cite{Eins},
Glashow-Weinberg-Salam (GWS) model for weak interactions \cite{Glas}, quantum
chromodynamics (QCD) for strong interactions \cite{Frit}, and grand unified theory
(GUT) \cite{Geor}.

Quantum gravity (QG) is introduced as an $SU(N)$ gauge theory with
the $\Theta$ vacuum term, which suggests that a certain group $G$
for gravitational interactions leads to a group $SU(3)_I \times
SU(3)_C$ for weak and strong interactions through dynamical
spontaneous symmetry breaking (DSSB) \cite{Roh2,Roh3}; the group chain is $G
\supset SU(3)_I \times SU(3)_C$ \cite{Roh1,Roh11}.
The Lagrangian density for QG \cite{Roh1,Roh11} is given by
\begin{equation}
{\cal L}_{QG} = - \frac{1}{2} Tr  G_{\mu \nu} G^{\mu \nu}
+ \sum_{i=1}  \bar \psi_i i \gamma^\mu D_\mu \psi_i  + \Theta \frac{g_g^2}{16 \pi^2} Tr G^{\mu \nu} \tilde G_{\mu \nu},
\end{equation}
where the bare $\Theta$ term \cite{Hoof2} is a nonperturbative term added to the
perturbative Lagrangian density with an $SU(N)$ gauge invariance. The subscript $i$
stands for the classes of pointlike spinor $\psi$ and $A_{\mu} = \sum_{a=0} A^a_{\mu}
\lambda^a /2$ stand for gauge fields. The field strength tensor is given by $G_{\mu
\nu} = \partial_\mu A_\nu - \partial_\nu A_\mu - i g_g [A_\mu, A_\nu]$ and $\tilde
G_{\mu \nu}$ is the dual field strength tensor. The $\Theta$ term apparently odd under
both P and T operation plays the role relating two different worlds, matter and
vacuum. The forms of Lagrangian densities are commonly analogous for gravitational,
weak, and strong interactions. Newton gravitation constant is defined as the effective
coupling constant
\begin{equation}
G_N/\sqrt{2}= g_f g_g^2/8 M_G^2 \approx 10^{-38} \ \textup{GeV}^{-2}
\end{equation}
with the gravitational coupling constant $g_g$, and the gauge boson mass $M_G \simeq
M_{Pl} \approx 10^{19}$ GeV in QG \cite{Roh1,Roh11} just as Fermi weak constant is
defined as $G_F/\sqrt{2}= g_w^2/8 M_W^2 \approx 10^{-5} \ \textup{GeV}^{-2}$ with the
weak coupling constant $g_w$ and the intermediate vector boson mass $M_W$ in weak
interactions \cite{Glas,Roh2}. The gauge boson mass $M_G$ decreases through the
condensation of the singlet graviton:
\begin{equation}
M_G^2 = M_{Pl}^2 - g_f g_g^2 \langle \phi \rangle^2 = g_f g_g^2 [A_{0}^2 -
\langle  \phi \rangle^2]
\end{equation}
where $A_{0}$ is the singlet gauge boson with even parity and $\langle  \phi \rangle$
is the condensation of singlet gauge boson with odd parity. QCD as an $SU(3)_C$ gauge
theory is the analogous dynamics of QWD as an $SU(3)_I$ gauge theory; QCD produces QND
as an $SU(2)_N \times U(1)_Z$ gauge theory for nuclear interactions and then produces
a $U(1)_f$ gauge theory for massless gauge boson (photon) dynamics just as QWD
produces the GWS model as an $SU(2)_L \times U(1)_Y$ gauge theory for weak
interactions and then produces a $U(1)_e$ gauge theory for photon dynamics
\cite{Roh2,Roh3}. The constraint of the extremely flat universe
\begin{math}
\Omega - 1 = - 10^{-61} ,
\end{math}
required by gauge theories \cite{Glas,Frit} and inflation scenario \cite{Guth} and
confirmed by the experiments BUMERANG-98 and MAXIMA-1 \cite{Jaff}, leads to
\begin{equation}
\Omega = (\langle \rho_m \rangle - \Theta \ \rho_m)/\rho_G = 1 - 10^{-61} ,
\end{equation}
where $\rho_m$ is the matter energy density, $\langle \rho_m \rangle$ is the zero
point energy density, and $\rho_G$ is the vacuum energy density. This means that the
ratio of the zero point energy density to the vacuum energy density is $\langle \rho_m
\rangle/\rho_G = 1$ and the $\Theta$ constant is obtained by
\begin{equation}
\Theta = 10^{-61} \ \rho_G/\rho_m .
\end{equation}
If the matter energy density in the universe $\rho_m \simeq \rho_c \simeq 10^{-47} \
\textup{GeV}^4$ is conserved, the $\Theta$ constant depends on the gauge boson mass
$M_G$ since $\rho_G = M_G^4$: $\Theta = 10^{-61} \ M_G^4/\rho_c$. $- 10^{-61}$
represents the $10^{30}$ expansion in one dimension, which is required by the
inflation scenario \cite{Guth}: $N_R = i/(\Omega -1)^{1/2} \approx 10^{30}$. The value
of $\Theta$ also makes the relation of the matter energy density $\rho_m$ to the
effective cosmological constant $\Lambda_e$: $\rho_m = \frac{\rho_G}{10^{61} \Theta} =
\frac{M_G^4}{10^{61} \Theta} = \frac{\Lambda_e}{10^{61} \Theta 8 \pi G_N}$. The
$\Theta$ value is connected with the effective cosmological constant defined by
\begin{equation}
\Lambda_e = 8 \pi G_N M_G^4
\end{equation}
where $\Lambda_{Pl} \approx 10^{122} \rho_c = 10^{38} \ \textup{GeV}^2$ at the Planck
epoch, $\Lambda_{EW} \approx 10^{-30}$ $\textup{GeV}^2$ at the weak epoch,
$\Lambda_{QCD} \approx 10^{-42}$ $\textup{GeV}^2$ at the strong epoch, and $\Lambda_0
= 3 H_0^2 \approx 10^{-84}$ $\textup{GeV}^2$ at the present epoch. In matter mass
generation of references \cite{Roh11,Roh2,Roh3}, the difference number of even-odd
parity singlet fermions $N_{sd}$ in intrinsic two-space dimensions suggests the a
degenerated particle number $N_{sp}$ in the intrinsic radial coordinate and an
intrinsic principal number $n_m$; these quantum numbers are connected by the relation
$n_m^4 = N_{sp}^2 = N_{sd}$ and the Dirac quantization condition $\sqrt{g_f} g_g
g_{gm} = 2 \pi N_{sp}$ is satisfied. The intrinsic principal quantum number $n_m$
consists of three quantum numbers, $n_m = (n_c, n_i, n_s)$ where $n_c$, $n_i$, and
$n_s$ are color, isospin, spin intrinsic principal quantum numbers respectively: total
intrinsic angular momentum is the vector sum of spin, isospin, and colorspin, $\vec
J_I = \vec S + \vec I + \vec C$.

In this approach \cite{Roh1,Roh11,Roh2,Roh3,Roh31}, vacuum energy and matter energy
are spatially quantized as well as photon energy and phonon energy. The vacuum
represented by massive gauge bosons is quantized by the maximum wavevector mode
\begin{equation}
N_R = i/(\Omega-1)^{1/2} \approx 10^{30}
\end{equation}
and the total gauge boson number $N_G =
4 \pi N_R^3/3 \approx 10^{91}$. The maximum wavevector mode $N_R
\approx 10^{30}$ is manifest since the universe size is $R_{Pl}
\approx 10^{-3}$ cm at the Planck scale $l_{Pl} \approx 10^{-33}$
cm and the universe size is $R_{0} \approx 10^{28}$ cm at the
present scale $l_{Pl} \approx 10^{-3}$ cm in the
extremely flat universe, $\Omega - 1 = - 10^{-61}$. Baryon matter
represented by massive baryons is quantized by the maximum
wavevector mode (Fermi mode) $N_F \approx 10^{26}$ and the total
baryon number $B = N_B = 4 \pi N_F^3/3 \approx 10^{78}$. This is
the result of the conservation of the baryon number. Baryon matter
quantization is consistent with the nuclear matter number density
\begin{equation}
n_n = n_B = A/(4 \pi r^3/3) \approx 1.95 \times 10^{38} \ \textup{cm}^{-3}
\end{equation}
given by the nuclear mass number $A = B$ and the
nuclear matter radius $r = r_0 A^{1/3}$ and with Avogadro's number
$N_A = 6.02 \times 10^{23} \ \textup{mol}^{-1} \approx 10^{19} \
\textup{cm}^{-3}$ in the matter. The baryon number density at the
nuclear interaction scale $10^{-1}$ GeV is $10^{26} \
\textup{cm}^{-3}$ in the universe size $R_{QCD} \approx 10^{17}$
cm, whose volume $10^{51} \ \textup{cm}^{3}$ is $10^{12}$ times
bigger than the matter volume $10^{39} \ \textup{cm}^{3}$.
Electrons with the mass $0.5$ MeV might be similarly quantized by
$N_F \approx 10^{27}$ and the total number $10^{81}$ if the
electron number is conserved as the result of the electron number
conservation under the assumption of $\Omega_e = \rho_e/\rho_c
\approx 1$. The maximum wavevector mode $N_F$ is close to
$10^{30}$ if the mass quantization unit of fermions $10^{-12}$ GeV
(the baryon number $B \simeq 10^{-12}$) is used rather than the
mass unit of baryons $0.94$ GeV ($B = 1$) under the assumption of
the fermion number conservation. Massless photons are quantized by
the maximum wavevector mode $N_\gamma \approx 10^{29}$ and the
total photon number $N_{t \gamma} = 4 \pi N_\gamma^3/3 \approx
10^{88}$. Cosmic microwave background radiation (CMBR) is the
conclusive evidence for massless gauge bosons (photons) with the
total number $N_{t \gamma} \approx 10^{88}$. Massless phonons in
the matter space are quantized by the maximum wavevector mode
(Debye mode) $N_D \approx 10^{25}$ and the total phonon number
$N_{t p} = 4 \pi N_D^3/3 \approx 10^{75}$.

Total particle numbers such as the gauge boson number $N_G \approx 10^{91}$, the
baryon number $N_B \approx 10^{78}$, the electron number $N_e \approx 10^{81}$, the
photon number $N_{t \gamma} \approx 10^{88}$, and the phonon number $N_{t p} \approx
10^{75}$ are conserved good quantum numbers as described above. The matter current at
the Planck scale, the (V - A) current and the electromagnetic current at the
electroweak scale, the baryon current and the proton current at the strong scale, and
the lepton current at the present scale might be conserved currents at different
energy scales. The proton number conservation is the consequence of the $U(1)_f$ gauge
theory just as the electron number conservation is the consequence of the $U(1)_e$
gauge theory. In gravitational interactions, the predicted typical lifetime for a
particle with the mass $1$ GeV is $\tau_p = 1/\Gamma_p \simeq 1/G_N^2 m^5 \approx
10^{50}$ years using the analogy of the lifetime of the muon $\tau_\mu = 192
\pi^3/G_F^2 m_\mu^5$ in weak interactions. Therefore, in the proton decay $p
\rightarrow \pi^0 + e^+$ at the energy $E << M_{Pl}$, the proton would have much
longer lifetime than $10^{32}$ years predicted by GUT \cite{Geor} if the decay process
is gravitational. In fact, the lower bound for the proton lifetime is $10^{32}$ years
at the moment. If the electric charge is completely conserved, the electron can not
decay. The present lower bounds for the electron lifetime are bigger than $10^{21}$
years for the electron decay into neutral particles and $10^{25}$ years for the decay
$e^{-} \rightarrow \gamma + \nu_e$ \cite{Stei}. The baryon number conservation is the
result of the $U(1)_Z$ gauge theory for strong interactions just as the lepton number
conservation is the result of the $U(1)_Y$ gauge theory for weak interactions. The
bound of the lepton number nonconservation process is expressed by the branching ratio
$B (K^+ \rightarrow \pi^- e^+e^+) < 10^{-8}$ or $B (\mu^-N \rightarrow e^+N') < 7
\times 10^{-11}$ and the bound of lepton flavor violation is shown by $B (\mu^-
\rightarrow e^- \gamma) < 5 \times 10^{-11}$, $B (\mu^- \rightarrow e^- \gamma \gamma)
< 7 \times 10^{-11}$, or $B (\mu^- \rightarrow 3 e^-) < 10^{-13}$. Table \ref{coga}
shows relations between conservation laws and gauge theories in strong and weak
interactions. Conservation laws of the baryon number, lepton number, and electric
charge number are good in weak, strong, and present interactions but they would be
nonperturbatively violated in gravitational interactions: there is possibility for
such violations even at much lower energy although they are extremely small. Discrete
symmetries such as parity (P), charge conjugate (C), time reversal (T), and charge
conjugate and parity (CP) are conserved perturbatively but are violated
nonperturbatively during DSSB. These violations are analogous to the nonconservation
of the (V + A) current in weak interactions and the axial vector current in strong
interactions. The breaking of discrete symmetries through the condensation of singlet
gauge bosons causes various asymmetries at different scales: the matter-antimatter
asymmetry with $\Theta_{Pl} \simeq 10^{61}$ at the Planck scale, the lepton-antilepton
asymmetry with $\Theta_{EW} \simeq 10^{-4}$ at the weak scale, and the
baryon-antibaryon asymmetry with $\Theta_{QCD} \simeq 10^{-12}$ at the strong scale.
The absence of the right-handed neutrino shows P violation \cite{Lee} and the decay of
the neutral kaon \cite{Chri} and the electric dipole moment of electrons \cite{Murt}
show CP violation by intermediate vector bosons. Pseudoscalar and vector mesons are
observable while their parity partners, scalar and pseudovector mesons, are not
observable because of massive gluons: the absence of the $U(1)_A$ particle is due to
the nonconservation of the color axial vector current.  Similarly, there are no baryon
octet and decuplet parity partners. The evidence of CP and T violation appears in the
electric dipole moment of the neutron \cite{Alta}. The conservation of the fermion
number $N_f \simeq 10^{91}$ in the unit of mass $10^{-12}$ GeV is good at the Planck
scale under the assumption with no supersymmetry and higher dimensions. Features of
conservation laws for fundamental forces are summarized in Table \ref{cola}.

The baryon asymmetry due to gravity is at present very weak by the
effective coupling constant $G_N \approx 10^{-38} \
\textup{GeV}^{-2}$. In terms of the baryon  energy density $\rho_B
\approx 1.88 \times 10^{-29} \ \Omega_B h_0^2 \ \textup{g \
cm}^{-3}$, the number of protons per unit volume is $n_B =
\rho_B/m_p \sim 1.13 \times 10^{-5} \ \Omega_B h_0^2 \
\textup{cm}^{-3}$. The baryon-antibaryon asymmetry at present is
estimated by the number of baryons dominating over the number of
antibaryon by a tiny factor of $10^{-10}$ if $\Omega_B \approx
0.1$:
\begin{equation}
\delta_B = \frac{N_B - N_{\bar B}}{N_B + N_{\bar B}}=
\frac{N_B}{N_{t \gamma}} \approx \frac{10^{78}}{10^{88}} = 10^{-10}
\end{equation}
where $N_{t \gamma}$ is the total number of massless gauge bosons (photons). The
lepton-antilepton asymmetry, which implies the lepton number violation observable at
present, is an analogue of the baryon asymmetry: $\delta_L = \frac{N_L - N_{\bar
L}}{N_L + N_{\bar L}} = \frac{N_L}{N_{t \gamma}}$ with the total lepton number $N_L =
L$. It consist of the electron asymmetry
\begin{equation}
\delta_e = \frac{N_e - N_{\bar e}}{N_{t \gamma}} \approx \frac{10^{81}}{10^{88}} =
10^{-7}
\end{equation}
and the neutrino asymmetry $\delta_\nu = \frac{N_\nu - N_{\bar \nu}}{N_{t \gamma}}
\approx \frac{10^{91}}{10^{88}} = 10^3$ according to the electron mass $0.5$ MeV and
the probable neutrino mass around $10^{-3}$ eV under the assumption of $\Omega_e =
\rho_e/\rho_c \approx 1$. The neutrino asymmetry is $\delta_\nu = N_\nu/N_{t \gamma}
\approx 10^{88}/10^{88} = 1$ if the neutrino mass $m_\nu \approx 1$ eV. Lepton
asymmetries of muon and tau $\delta_\mu = N_\mu/N_{t \gamma} \approx 10^{79}/10^{88} =
10^{-9}$ and $\delta_\tau = N_\tau/N_{t \gamma} \approx 10^{78}/10^{88} = 10^{-10}$
are as well expected if lepton matter has the same order with the critical density
$\rho_c$. The baryogenesis and leptogenesis described above can be discussed from the
gauge theory point of view. Massive gravitons might violate discrete symmetries and
the antimatter (or antibaryon or antilepton) number conservation just as the Higgs
mechanism in electroweak interactions violates discrete symmetries and chiral
symmetry. The discrete symmetries of P, C, CP, and T are explicitly broken during DSSB
as the requirement of the baryon asymmetry. If the antimatter number current is not
conserved, some antimatter particle spectra must disappear. Since the interaction rate
is given by $\Gamma \sim n \sigma |v| \sim G_N^2 T^5$ and the expansion rate is given
by $H_e \sim T^2/M_{Pl}$, the ratio of the interaction rate to the expansion rate
becomes $\Gamma/H_e \sim T^3/M_{Pl}^3$, which indicates nonequilibrium starts at the
temperature $T\sim M_{Pl}$. The population of gravitons and the number of antimatter
were suppressed by the Boltzmann factor $\textup{exp} (-M_{Pl}/T)$ since $T < M_{Pl}$.
Baryon and lepton asymmetries predicted under the assumption of $\Omega_B =
\rho_B/\rho_c \approx 1$ and $\Omega_L = \rho_L/\rho_c \approx 1$ seem to indicate the
nonconservation of the lepton and baryon quantum numbers separately above the weak
scale since $\delta_L \neq \delta_B$. The baryon number is conserved below the strong
scale as illustrated by the $U(1)_Z$ gauge theory and the lepton number is conserved
below the weak scale as illustrated by the $U(1)_Y$ gauge theory but they may not be
separately conserved and only the $(B - L)$ quantum numbers may be conserved above the
weak scale. In the minimal GUT of the $SU(5)$ gauge theory \cite{Geor}, the matter
asymmetry term involved in perturbation theory is too small to explain the observed
baryon asymmetry. However, nonperturbative processes during DSSB by QG as a gauge
theory indicates that the asymmetries are carried by the current anomaly and the gauge
boson condensation.

Table \ref{fuco} summarizes fundamental and cosmological constants in quantum
cosmology \cite{Roh11}: the gauge boson mass $M_G$, the effective coupling constant
$G_G \simeq \sqrt{2} g^2/8 M_G^2$, the gauge boson number density $n_G \simeq M_G^3$,
the vacuum energy density $V_e (\bar \phi) \simeq M_G^4$, the cosmological constant
$\Lambda_e \simeq 8 \pi G_N M_G^4$, the Hubble constant $H_e = (\Lambda_e/3)^{1/2}
\simeq (8 \pi G_N M_G^4/3)^{1/2}$, the baryon number density $n_B$, the baryon mass
density $\rho_B$, the electron number density $n_e$, the electron mass density
$\rho_e$, the photon energy $E_\gamma$, the photon number density $n_{t \gamma} \simeq
2 \zeta(3) T^3/\pi^2$, the photon energy density $\epsilon_\gamma$, the phonon number
density $n_{t p}$, the phonon energy density $\epsilon_p$, the $\Theta$ constant
$\Theta \simeq 10^{-61} \rho_G/\rho_m$, and the intrinsic topological constant $\nu
\simeq \rho_m/\rho_G$. The values of the baryon (electron) number density and mass
density represent ones when the vacuum volume is used while the values within
parentheses represent ones when only the baryon (electron) matter volume is used.

New noble concepts based on local gauge theories for fundamental forces are introduced
under the constraint of the flat universe, which is required by quantum gauge theories and
is confirmed by the recent experiments BUMERANG-98 and MAXIMA-1.
Intrinsic quantum numbers emerge in analogy
with extrinsic quantum numbers. Fundamental constants and cosmological parameters are
qualitatively and quantitatively discussed as the consequences of gauge invariance for
fundamental forces, which lead to conservation laws. The proton number conservation is
the consequence of the $U(1)_f$ gauge theory just as the electron number conservation
is the consequence of the $U(1)_e$ gauge theory and the baryon number conservation is
the result of the $U(1)_Z$ gauge theory for strong interactions just as the lepton
number conservation is the result of the $U(1)_Y$ gauge theory for weak interactions.
However, there is possibility not for the separate conservation of the baryon number
or the lepton number but for the combined conservation of the $(B - L)$ number
conservation in the energy regions above $10^2$ GeV. Discrete symmetries C, P, T, and
CP are nonperturbatively violated during DSSB although they are not perturbatively
violated. The apparent baryon asymmetry indicates the nonperturbative violation from
the gauge theory point of view for gravitation and the possible lepton asymmetry
emerges in analogy with the baryon asymmetry.  Experimental precision tests such as the proton
decay, lepton flavor violation, electric charge violation, etc. as well as
nonperturbative discrete symmetry violation are demanded at higher and lower energies.
This work may thus significantly contribute to the understanding of profound
underlying principles and the unification of fundamental forces since fundamental
constants and conservation laws are remarkably clarified.

\newpage
\onecolumn

\begin{table}
\caption{\label{coga} Relations between Conservation Laws and
Gauge Theories}
\end{table}
\centerline{
\begin{tabular}{|c|c|c|} \hline
Force & Conservation Law & Gauge Theory \\ \hline \hline
Electromagnetic & Proton & $U(1)_f$ \\ \hline
Strong & Baryon & $U(1)_Z$ \\ \hline
Strong & Color Vector & $SU(2)_N \times U(1)_Z$ \\ \hline
Strong & Color & $SU(3)_C$ \\ \hline
Electromagnetic & Electron & $U(1)_e$ \\ \hline
Weak & Lepton & $U(1)_Y$ \\ \hline
Weak & V-A & $SU(2)_L \times U(1)_Y$ \\ \hline
Weak & Isotope (Isospin) & $SU(3)_I$ \\ \hline
\end{tabular}
}

\vspace{1cm}

\begin{table}
\caption{\label{cola} Overview of Conservation Laws}
\end{table}
\centerline{
\begin{tabular}{|c|c|c|c|c|} \hline
Conservation  & Gravity & Electromagnetic & Weak & Strong \\ \hline \hline Energy,
Momentum, Angular Momentum & yes & yes & yes & yes \\ \hline Charge, Baryon, Lepton &
no & yes & yes & yes \\ \hline P, C, T, CP & no & yes & no & no \\ \hline TCP & yes &
yes & yes & yes \\ \hline
\end{tabular}
}

\vspace{1cm}

\begin{table}
\caption{\label{fuco} Fundamental and Cosmological Constants in
Quantum Cosmology}
\end{table}
\centerline{
\begin{tabular}{|c|c|c|c|c|} \hline
Constant  & Gravity  & Weak & Strong & Present \\ \hline \hline
Gauge Boson Mass $M_G$ (GeV) & $10^{19}$ & $10^2$ & $10^{-1}$ & $10^{-12}$ \\ \hline
Effective Coupling Constant $G_G$ ($\textup{GeV}^{-2}$) & $10^{-38}$ &  $10^{-5}$
& $10^{-1}$ & $10^{24}$ \\ \hline
Gauge Boson Number Density $n_G$ ($\textup{cm}^{-3}$) & $10^{98}$ & $10^{47}$ & $10^{39}$ & $10^{5}$ \\ \hline
Vacuum Energy Density $V_e$ ($\textup{g} \ \textup{cm}^{-3}$) & $10^{93}$ &  $10^{25}$ & $10^{14}$ & $10^{-29}$ \\ \hline
Cosmological Constant $\Lambda_e$ ($\textup{GeV}^2$) & $10^{38}$ & $10^{-30}$ & $10^{-42}$ & $10^{-84}$ \\ \hline
Hubble Constant $H_e$ ($\textup{GeV}$) & $10^{19}$ & $10^{-15}$ & $10^{-21}$ & $10^{-42}$  \\ \hline
Baryon Number Density $n_B$ ($\textup{cm}^{-3}$) & & & $10^{26}(10^{38})$ & $10^{-6}(10^{4})$ \\ \hline
Baryon Mass Density $\rho_B$ ($\textup{g} \ \textup{cm}^{-3}$) & & & $10^{1}(10^{14})$ & $10^{-31}(10^{-20})$ \\ \hline
Electron Number Density $n_e$ ($\textup{cm}^{-3}$) & & $10^{38}(10^{49})$ & $10^{29}(10^{41})$ & $10^{-3}(10^{7})$ \\ \hline
Electron Mass Density $\rho_e$ ($\textup{g} \ \textup{cm}^{-3}$) & & $10^{23}(10^{35})$ & $10^{1}(10^{14})$ & $10^{-31}(10^{-20})$ \\ \hline
Photon Energy $E_\gamma$ ($\textup{GeV}$) & $10^{18}$ & $10^{1}$ & $10^{-2}$ & $10^{-13}$ \\ \hline
Photon Number Density $n_{t \gamma}$ ($\textup{cm}^{-3}$) & $10^{95}$ & $10^{44}$ & $10^{36}$ & $10^{2}$ \\ \hline
Photon Energy Density $\epsilon_\gamma$ ($\textup{g} \ \textup{cm}^{-3}$) &
$10^{89}$ & $10^{21}$ & $10^{10}$ & $10^{-34}$ \\ \hline
Phonon Number Density $n_{t p}$ ($\textup{cm}^{-3}$) & & & $10^{24}$ & $10^{-10}$ \\ \hline
Phonon Energy Density $\epsilon_p$ ($\textup{g} \ \textup{cm}^{-3}$) & & & $10^{-2}$ & $10^{-46}$ \\ \hline
Constant $\Theta$ & $10^{61}$ & $10^{-4}$ & $10^{-12}$ & $10^{-61}$ \\ \hline
Topological Constant $\nu$ & $10^{-122}$ & $10^{-57}$ & $10^{-49}$ & $10^{0}$ \\ \hline
\end{tabular}
}

\end{document}